\documentclass[screen,sigconf]{acmart}
\usepackage{amsmath}
\usepackage{array,booktabs}
\usepackage{xcolor}
\usepackage{euler}
\usepackage[T1]{fontenc}
\usepackage{wasysym}
\usepackage{xspace}
\usepackage{array}
\usepackage{multicol}        
\usepackage{multirow}        
\usepackage{natbib}
\usepackage{algorithm}
\usepackage[]{algpseudocode}
\usepackage{todonotes}

\newtheorem{theorem}{Theorem}

\newtheorem{definition}[theorem]{Definition}

\newtheorem{remark}[theorem]{Remark}

\newcommand{\ffour}{\texttt{F4}\xspace}

\newcommand{\fglm}{\texttt{fglm}\xspace}

\newcommand{\msolve}{\texttt{msolve}\xspace}
\newcommand{\fgb}{\texttt{FGb}\xspace}
\newcommand{\slv}{\texttt{SLV}\xspace}
\newcommand{\tdescartes}{\texttt{tdescartes}\xspace}
\newcommand{\magma}{\texttt{magma}\xspace}
\newcommand{\maple}{\texttt{maple}\xspace}

\newcommand{\gmp}{\textsc{GMP}\xspace}
\newcommand{\gc}{Gr\"ob\-ner con\-ver\-sion\xspace}
\newcommand{\gb}{Gr\"ob\-ner ba\-sis\xspace}
\newcommand{\gbs}{Gr\"ob\-ner ba\-ses\xspace}

\newcommand{\field}{\ensuremath{\mathcal{K}}\xspace}
\newcommand{\afield}{\ensuremath{\overline{\field}}\xspace}
\newcommand{\pr}{\ensuremath{\mathcal{P}}\xspace}

\newcommand{\N}{\ensuremath{\mathbb{N}}\xspace}
\newcommand{\Q}{\ensuremath{\mathbb{Q}}\xspace}
\newcommand{\Z}{\ensuremath{\mathbb{Z}}\xspace}

\DeclareMathOperator{\DRL}{DRL}
\DeclareMathOperator{\LEX}{LEX}
\newcommand{\sdrl}{\ensuremath{\mathrel{<_{\DRL}}}\xspace}

\newcommand{\slex}{\ensuremath{\mathrel{<_{\LEX}}}\xspace}

\DeclareMathOperator{\lcm}{lcm}

\DeclareMathOperator{\lcop}{lc}
\newcommand{\lc}[2]{\lcop_{#2}\left({#1}\right)}
\DeclareMathOperator{\lmop}{lm}
\newcommand{\lm}[2]{\lmop_{#2}\left({#1}\right)}
\DeclareMathOperator{\ltop}{lt}
\newcommand{\lt}[2]{\ltop_{#2}\left({#1}\right)}
\DeclareMathOperator{\spairop}{sp}
\newcommand{\spair}[2]{\spairop\left({#1,#2}\right)}

\newcommand{\assumesc}{(\textsf{\texorpdfstring{P\textsubscript{1}}{P1}})\xspace}
\newcommand{\assumesl}{(\textsf{\texorpdfstring{P\textsubscript{2}}{P2}})\xspace}
\newcommand{\up}{f}
\newcommand{\mulmat}{\mathcal{M}}
\newcommand{\mbasis}{B}
\newcommand*{\defeq}{\coloneqq}

\DeclareMathOperator\quo{quo}

\DeclareMathOperator\Berlekamp{Berlekamp--Massey}
\DeclareMathOperator\squarefree{squarefree}
\newcommand{\elim}{\ensuremath{g_n}\xspace}
\newcommand{\elimsqfr}{\ensuremath{w_n}\xspace}
\newcommand{\param}[1]{\ensuremath{f_{#1}}\xspace}
\newcommand{\paramsqfr}[1]{\ensuremath{u_{#1}}\xspace}
\newcommand{\vvec}{\ensuremath{V}\xspace}
\newcommand{\resp}{resp.\xspace}
\newcommand{\wrt}{w.r.t.\xspace}
\newcommand{\eg}{e.g.\xspace}
\newcommand{\ie}{i.e.\xspace}
\newcommand\revision[1]{#1}

\author{J\'er\'emy Berthomieu}
\affiliation{%
	\institution{Sorbonne Universit\'e, \textsc{CNRS}, \textsc{LIP6}}
	\city{F-75005 Paris}
	\postcode{75252}\country{France}}
\email{jeremy.berthomieu@lip6.fr}

\author{Christian Eder}
\affiliation{%
	\institution{Technische Universit\"at Kaiserslautern}
	\streetaddress{Gottlieb-Daimler-Str.}
	\city{Kaiserslautern}
	\postcode{67663}\country{Germany}}
\email{ederc@mathematik.uni-kl.de}

\author{Mohab Safey El Din}
\affiliation{%
	\institution{Sorbonne Universit\'e, \textsc{CNRS}, \textsc{LIP6}}
	\city{F-75005 Paris}
	\postcode{75252}\country{France}}
\email{mohab.safey@lip6.fr}

\title{
\texttt{msolve}: A Library for Solving Polynomial Systems}

\copyrightyear{2021}
\acmYear{2021}
\setcopyright{acmcopyright}\acmConference[ISSAC '21]{Proceedings of the
  2021 International Symposium on Symbolic and Algebraic
  Computation}{July 18--23, 2021}{Virtual Event, Russian Federation}
\acmBooktitle{Proceedings of the 2021 International Symposium on
  Symbolic and Algebraic Computation (ISSAC '21), July 18--23, 2021,
  Virtual Event, Russian Federation}
\acmPrice{15.00}
\acmDOI{10.1145/3452143.3465545}
\acmISBN{978-1-4503-8382-0/21/07}

\begin{document}
\begin{abstract}
  We present a new open source C library \texttt{msolve} 
  dedicated to solving multivariate
  polynomial systems of dimension zero through computer algebra methods. 
  The core algorithmic framework of \texttt{msolve} relies on Gr\"obner bases
  and linear algebra based algorithms for polynomial system solving. 
  It relies on Gr\"obner basis computation w.r.t.\ the degree reverse lexicographical
  order, Gr\"obner conversion to a lexicographical Gr\"obner basis
  and real solving of univariate  
  polynomials.
  We explain in detail how these three main steps
  of the solving process are implemented, how we exploit \texttt{AVX2} instruction
  processors and the more general implementation ideas we put into practice
  to better exploit the computational capabilities of this algorithmic
  framework. 
  We compare the practical performances of \texttt{msolve} with 
  leading computer
  algebra systems such as \textsc{Magma}, \textsc{Maple},
  \textsc{Singular} on a wide range of  
  systems with finitely many complex solutions, showing that
  \texttt{msolve} can tackle systems which were out of reach by the
  computer algebra software state-of-the-art. 
\end{abstract}
\thanks{%
  The first and third authors are supported by the joint ANR-FWF
  ANR-19-CE48-0015 \textsc{ECARP} project, the ANR grants ANR-18-CE33-0011
  \textsc{Sesame} and ANR-19-CE40-0018 \textsc{De Rerum Natura}
  projects, the PGMO grant \textsc{CAMiSAdo} and the European
  Union's Horizon 2020 research and innovation programme under the Marie
  Sklodowska-Curie grant agreement N. 813211 (POEMA). The second author is
  supported by the Forschungsinitiative Rheinland-Pfalz.}
\maketitle

\vspace*{-1em}
\section{Introduction}
\label{sec:intro}

\paragraph*{Problem statements and motivation}
Polynomial systems arise in a wide range of areas of scientific engineering and
computing sciences. Classical problems are to decide if the solution set is
finite (over an algebraic closure of the ground field), compute its dimension
when it is not, else count the solutions and isolate them over the real or
complex numbers when the ground field is infinite. 


We design a software library, for solving multivariate polynomial systems, with
a focus on those which have dimension at most $0$, i.e. finitely many solutions
in an algebraic closure of the ground field. We rely on computer algebra methods
yielding algebraic parametrizations of their solutions. This allows us to bypass
the commonly encountered issues related to accuracy and exhaustivity met by
numerical methods because of the non-linearity of the input.


\paragraph*{Prior works and state-of-the-art}  In this context, one can 
mention \emph{regular chains} whose base operation is computing gcd of 
polynomials with coefficients encoded by an algebraic tower of extensions 
combined with splitting polynomial ideal techniques~\cite{ALM}, 
\emph{geometric resolutions} which is based on an incremental procedure 
intersecting a lifted curve (obtained by Hensel lifting generic solutions 
to the first $i$ polynomials) with the hypersurface defined by the $(i+1)$st 
polynomial~\cite{GiustiLS2001}
and \emph{Gr\"obner bases} which consist in 
a set of polynomials in the ideal generated by the input such that for a 
given monomial order one can use them to define an intrinsic multivariate 
division and
thus
decide the ideal 
membership problem. 

In \msolve, we focus on \gbs because of their importance in computer 
algebra systems and their use in many higher-level algorithms. Note that 
when the input system generates a radical ideal, of dimension at most $0$, 
and in generic coordinates, a \gb for a lexicographical order on the 
monomials is in a so-called \emph{shape position}, \ie it has the 
following shape:
\begin{equation}
  w(x_n), x_{n-1} + u_{n-1}(x_n), \ldots, x_1 +
  u_1(x_n).\label{eq:radparam}
\end{equation}
One can then recover the coordinates of all solutions by evaluating univariate 
polynomials at the roots of a univariate polynomial. Up to normalization, 
this is very close to a rational parametrization
\begin{equation}
  w(x_n), w'(x_n) x_{n-1} + v_{n-1}(x_n), \ldots, w'(x_n)x_1 + v_1(x_n)
  \label{eq:rational}
\end{equation}
where $w'$ is the derivative of $w$. Such representations of the solution set go
back to Kronecker and appear in many works (see \eg~\cite{AlonsoBRW1996,
  Kronecker}) and, under the above assumptions, are computed by regular chains
and geometric resolution algorithms.

\revision{Further, we mean by solving a polynomial system of dimension at most
  $0$ the computation of such a rational parametrization of its solution set.
  Note that such a parametrization exists only when all distinct solutions have
  distinct $x_n$-coordinate, 
  which one can always ensure through some linear change of coordinates. Also, when
  the input coefficients are rational numbers, one includes in the requested
  output the isolation of the real solutions to $w$. }

Several libraries for computing \gbs can be found, most of them being either
tailored for crypto applications (see \eg~\cite{M4RI})
or are designed for algorithmic experimentation (see \eg~\cite{tinygb}).
Recently, 
\maple and \magma have greatly improved their \gbs engines, the one in \maple being based
several years 
ago on \fgb~\cite{FGb}, which is developed by J.-Ch.~Faug\`ere.

\paragraph*{Main results} This is a software paper and hence does not 
contain any new algorithm or theorem. The main outcome of this work is a 
software library, written in plain \texttt{C}, \emph{open source}, distributed under 
the license \texttt{GPLv2} which includes modern implementations of 
algorithms for solving multivariate polynomial systems based on \gbs. It 
supports polynomial systems with coefficients in a prime field of 
characteristic $< 2^{31}$ or with rational coefficients. It allows one to 
solve zero-dimensional systems which are out of reach for leading computer 
algebra systems like \magma and \maple.

For instance, \msolve is able to solve polynomial systems with thousands of 
complex solutions such as Katsura-14, which has $8,192$ complex solutions, 
whose solution set is encoded by a rational parametrization of bit size 
$\simeq 2^{32.37}$, sequentially, within $15$ days on an
\textsc{Intel\textsuperscript{\tiny\textregistered}}
\textsc{Xeon\textsuperscript{\tiny\textregistered}} 
\textsc{CPU E7-4820} v4 @ 2.00GHz while \maple and \magma could not solve it after 
$6$ months.

The \msolve library is available at: \hfill \url{https://msolve.lip6.fr}

It includes efficient implementations of the \ffour algorithm~\cite{F4} 
(reducing \gbs computations to Gaussian elimination), of a change of 
orders algorithm due to Faug\`ere and Mou~\cite{sparse-FGLM} (based on 
computing minimal polynomials of linear endomorphisms) and a dedicated 
univariate real root solver.  Solving systems with rational coefficients is 
handled through multi-modular computations. 

We design and use dedicated data structures to take into account current 
hardware architectures. Our \ffour implementation enjoys several 
implementations of linear algebra with dedicated storage to handle sparsity 
structures arising naturally in this algorithm as well as hashing tables 
and masks for encoding exponent vectors and divisibility checks between 
monomials.  

Our implementation of change of orders is designed for the cases where 
the \emph{radical} of the ideal generated by the input equations admits a 
\gb for a lexicographical order which is in shape position and under an 
extra assumption which is recovered by replacing the last variable by a 
generic enough linear combination of the input variables. Hence, \msolve 
loses information on the {multiplicities} of the solutions but focuses on 
solving.  It includes an efficient routine verifying the correctness of the 
result when the input ideal is not radical which was missing in the 
literature.

This implementation uses a dedicated storage of the matrix encoding the 
linear endomorphism for which one needs to compute the minimal polynomial.  
It exploits the structure of this matrix to reduce this computation to 
scalar products of dense vectors. 

This allows us to use extensively vectorization instructions such as 
\texttt{AVX2} to speed up our computations. A more intricate use of 
\texttt{AVX2} instructions is also set for the $\ffour$ implementation of 
\msolve. 

A special care has been brought to memory consumption which is low compared 
to the one of \maple or \magma. This is suitable for a trivial 
multi-threaded scheme for multi-modular computations, almost dividing the 
runtime by the number of threads. 

\paragraph*{Structure of the paper} In Section~\ref{sec:basics}, we fix 
some notation and recall some background.
Section~\ref{sec:algorithms} gives an overview of the algorithms.
Section~\ref{sec:implementation} describes the design of the library and 
the implementation ideas.
Section~\ref{sec:results} reports on the practical performances of 
\msolve. 


\section{Notations and Background}
\label{sec:basics}
We recall below some basic notions on polynomial rings and Gr\"obner bases.  
Let $\field$ be a field; we 
denote by $\pr = \field[x_1, \ldots, x_n]$ the polynomial ring with base 
field $\field$ and variables  $x_1, \ldots, x_n$.


Let $<$ denote a monomial order on \pr. We consider only \emph{global}
monomial orders, \ie orders for which $x_i > 1$ for all $1 \leq i \leq n$.
We mainly consider the global monomial orders $\sdrl$, the degree reverse lexicographical order, and
$\slex$, the lexicographical order (see \eg~\cite[Chap.~2, Sec.~2,
Def.~3]{CLO}).
Given a monomial order $<$ we can highlight the maximal terms of
elements in \pr with respect to $<$: For $f\in \pr\ \backslash\ \{0\}$,
${\lt{f}{<}}$ is the \emph{lead term}, $\lm{f}{<}$ the \emph{lead 
monomial}, and $\lc{f}{<}$
the \emph{lead coefficient} of $f$. For any set $F \subset \pr$ we define
the \emph{lead ideal} $L_{<}(F) = \langle \lt{f}{<} \mid f \in F\rangle$; 
for an ideal
$I \subset \pr$,\  $L_{<}(I)$ is defined as the ideal generated by lead 
terms of all elements of
$I$. Further, we omit the index $<$ when it is clear from the context.

\begin{definition}
\label{def:gb}
A finite set $G \subset \pr$ is called a \emph{\gb} for an ideal $I \subset \pr$
\wrt a monomial order $<$ if $G \subset I$ and $L_<(G) = L_<(I)$. 
This is equivalent to the condition that for any $f \in
I\backslash\{0\}$ there exists a $g\in G$ such that 
$\lm{g}{<} \mid \lm{f}{<}$. \end{definition}

Buchberger gave in $1965$~\cite{bGroebner1965} an algorithmic criterion for
computing \gbs based on the definition of S-pairs:

\begin{definition}
    Let $f,g\in \pr$ be nonzero, let $G \subset \pr$ be finite.
\begin{enumerate}
\item Denote by $\lambda \defeq
    \lcm\left(\lm{f}{<},\lm{g}{<}\right)$. The
\emph{S-pair} between $f$ and $g$ is given by
\[\spair{f}{g} \defeq \frac{\lambda}{\lt{f}{<}} f -
\frac{\lambda}{\lt{g}{<}}g.\]
\item We say that $g$ is a {reducer} of $f$ if for a term $t$ in $f$ there
    exists a term $\sigma \in \pr$ such that $\lt{\sigma g}{<} = t$. The
    \emph{reduction} of $f$ by $g$ is then given by $f-\sigma g$.
    We say that $f$ \emph{reduces to $h$ \wrt $G$ and $<$} if there exist
    finitely many reducers $g_1,\ldots,g_k$ in $G$
    such that $h = f - \sum_{i=1}^k \sigma_i g_i$ and there exists no term
    in $h$ for which an element of $G$ is a reducer.
    \end{enumerate}

    \label{def:spair}
\end{definition}

\begin{theorem}[Buchberger's criterion]
\label{thm:gb}
A finite set $G \subset \pr$ is called a \emph{\gb} for an ideal $I \subset \pr$ \wrt
a monomial order $<$ if for all $f,g \in G\; \spair{f}{g}$ reduces to zero \wrt $G$.
\end{theorem}

Gr\"obner bases for an ideal $I$ \wrt a monomial order $<$ are not 
unique, but \emph{reduced} Gr\"obner bases are (these are Gr\"obner bases 
where the lead coefficient of each element is $1$ and where no monomial of 
an element $g\in G$ lies in $L_{<}(G\setminus \{g\} )$).   

Recall that Gr\"obner bases allow one to define a \emph{normal form}, \ie  
given $f\in \pr$, $F\subset \pr$ and a Gr\"obner basis $G$ of  $\left<F 
\right>$ for a monomial order $<$, one can compute a unique representative 
of $f$ in the quotient ring  $\frac{\pr}{\left<F \right>}$ which is a 
$\field$-vector space. 
This property of a \gb allows one to discover 
\emph{relations}, \ie polynomials which lie in $\left< F \right>$ and the 
Hilbert series/polynomial associated to $\left<F \right>$
(see~\cite[Chap.~10, Sec.~2]{CLO}) 
when using the $\sdrl$ order, from which we deduce the Krull dimension and 
the degree of $\left<F \right>$. Recall that the Krull dimension of 
$\left<F \right>$ coincides with the dimension of its associated algebraic 
set $V(F)$ in $\afield^n$, \ie the largest integer $d$ such that the 
intersection of $V(F)$ with $d$ hyperplanes in $\afield^n$ is finite and of 
maximal  cardinality. This cardinality is the degree of $\left<F \right>$ 
when it is radical (\ie $f^k\in \left<F \right>$ for some $k$ implies that  
$f\in \left<F \right>$). We refer to~\cite[Chapter~12]{Eisenbud} for more details on 
the equivalence of the various definitions of dimension
and~\cite[Section~1.9]{Eisenbud} for 
the relations between Hilbert series and degree of ideals and varieties.

Elimination orders such as $\slex$ allow one to compute a basis for 
$\left<F \right>\cap \field[x_i, \ldots, x_n]$ for $1\le  i\le n$ and then 
put into practice the Elimination theorem~\cite[Chapter~3]{CLO}
by yielding 
an algebraic description of the Zariski closure of the projection of  
$V(F)$ on the $(x_i, \ldots, x_n)$-subspace. It turns out that Gr\"obner 
bases \wrt $\slex$ order enjoy a triangular structure from which one can 
extract information on the solution set.

Other geometric operations are encoded with ideal theoretic operations such 
as set difference whose algebraic counterpart is \emph{saturation} (given 
an ideal $I\subset \pr$ and  $g \in \pr$, the saturation of  $I$ by 
$\left<g \right>$, denoted by  $I:g^\infty$, is the set of polynomials $h$ 
such that  $hg^k\in I$ for some $k\in \N$). 

When the set of solutions of $F$ in  $\afield^n$ is finite, the quotient 
ring  $\frac{\pr}{\left<F \right>}$ is a finite dimensional vector
space~\cite[Chap.~5, Sec.~2, Prop.~7]{CLO}.
The dimension of this vector space coincides with the 
\emph{degree} of the ideal $\left<F \right>$: this is the 
number of solutions counted with multiplicities.

This property is at the foundations of algorithms based on Gr\"obner bases 
for solving polynomial systems. It implies that $\left<F \right>$ has a 
non-zero intersection with $\field[x_i]$ for  $1 \le i \le n$. It is 
heavily exploited in change of orders algorithms computing a Gr\"obner 
basis for a given ideal $\left<F \right>$, in particular to compute 
rational parametrizations as they are defined in Section~\ref{sec:intro}.  
Such algorithms are important since computing Gr\"obner bases \wrt  
$\slex$ order is usually way more expensive than pre-computing a Gr\"obner 
basis \wrt  $\sdrl$ and then applying such a change of orders 
algorithm~\cite{BayerStillman87}.   

In the following, we assume we are given a finite set of polynomials 
$F\subset \pr$ such that $\left<F \right>$ is zero-dimensional, that
is $F=0$ has
a finite number of solutions in $\afield^n$.  Hence, when $F$ 
satisfies the Shape position assumption described in 
Section~\ref{sec:intro}, we apply the following classical solving 
strategy:\\
(1) Compute the reduced \gb  $G$ of $\left<F \right>$ \wrt \sdrl.\\
    Note that once we have computed $G$ we can decide whether  $\left<F 
    \right>$ is zero-dimensional.  \\
(2) Convert $G$ to the reduced \gb $H$ of $\sqrt{ \left<F \right>}$ \wrt  
\slex and deduce a rational parametrization $R$ encoding its
solutions,
as in Equation~\eqref{eq:rational}.\\
(3) Apply a univariate solver to the uniquely defined univariate polynomial 
$w$ in $R$. Go on by substituting variables already solved. 


\vspace*{-1em}
\section{Implemented Algorithms}
\label{sec:algorithms}
\subsection{Faug\`ere's \ffour Algorithm}
\label{subsec:f4}
In $1965$, Buchberger initiated the theory
of \gbs for global monomial orders.
Specifically, he introduced some key
structural theory, and based on
this theory, proposed the first algorithm for computing \gbs~\cite{bGroebner1965, bGroebner1965eng}.
Buchberger's algorithm introduced the concept of critical pairs and
S-pairs and repeatedly carries out a certain
polynomial operation (called reduction).

\msolve includes an implementation of Faug\`ere's \ffour
algorithm~\cite{F4} which is variant of Buchberger's.
Here we highlight the main
differences to Buchberger's algorithm:

%
\noindent
$\bullet$ In contrast to Buchberger's algorithm one can choose
several S-pairs from the pair set $P$ at a time, for example, all of the same minimal degree. These
S-pairs are stored in a subset $L \subset P$.\\
$\bullet$ Then for all terms of all the generators 
of the S-pairs in $L$ we
    search in the current intermediate \gb $G$ for possible 
    reducers.
    We add those
to $L$ and again search all of their terms for reducers in $G$.\\
$\bullet$ Once all available reduction data is collected from the last 
step, we
generate a matrix with columns corresponding to the terms appearing in $L$ and
rows corresponding to the coefficients of each polynomial in $L$. In order to
reduce now all chosen S-pairs at once we apply Gaussian Elimination
on the matrix and recheck afterwards which rows of the
updated matrix give a new leading monomial not already in $L(G)$.

In order to optimize the algorithm one can now apply Buchberger's product and
chain criteria, see~\cite{bGroebnerCriterion1979, kollreider-buchberger-1978}.
With these, useless S-pairs are removed before even added to $P$ thus less zero
rows are computed during the linear algebra part of \ffour. Still, for bigger
examples there are many zero reductions.

It is known that Buchberger-like algorithms for computing \gbs, as \ffour, have
a worst-case time complexity doubly-expo\-nential in the number of solutions of the
system for $\sdrl$. Still, in practice these algorithms behave in
general way better.

\vspace*{-1em}
\subsection{\gc Algorithm}
\label{subsec:fglm}





We present the variant of the \fglm
algorithm~\cite{FGLM} due to
Faug\`ere and Mou~\cite{sparse-FGLM} which is used in \msolve. We assume 
that the input \gb $G$
is reduced and that it spans a
zero-dimensional ideal $I$ of degree $D$. We also assume that $I$
satisfies two \emph{generic} assumptions, namely assumptions~\assumesc
and~\assumesl defined below.

Assumption~\assumesc
means that for every
monomial $m$ in the monomial basis $\mbasis$ of $\frac{\pr}{I}$ for $\sdrl$,
either $m x_n$ is another monomial in $\mbasis$ or it is the
leading monomial of a polynomial in $G$.

Since $\frac{\pr}{I}$ is a finite dimensional $\field$-algebra, the
multiplication by $x_n$ is a linear map whose associated matrix
$\mulmat$ is
called the \emph{multiplication matrix of $x_n$}. Under
assumption~\assumesc, each column of $\mulmat$ is either a column of the 
identity matrix or
can be read from a polynomial in
$G$ whose leading term is divisible by $x_n$.

Assumption~\assumesl means that $I$ is in
\emph{shape position}, \ie the reduced \gb for
$I$ for $\slex$ is given by
$\{\elim(x_n),x_{n-1}+\param{n-1}(x_n),\ldots,x_1+\param{1}(x_n)\}$
where $\elim$ has degree $D$ and polynomials
$\param{1},\ldots,\param{n-1}$, the \emph{parametrizations} of
$x_1,\ldots,x_{n-1}$ have degrees at most
$D-1$. Furthermore, the radical of $I$, $\sqrt{I}=\{h\in\pr|\exists k\in\N,
h^k\in I\}$, also satisfies
\assumesl
and
its reduced \gb
for $\slex$ is
$\{\elimsqfr(x_n),x_{n-1}+\paramsqfr{n-1}(x_n),\ldots,x_1+\paramsqfr{1}(x_n)\}$,
with $\elimsqfr$ the squarefree part of $\elim$ and
$\deg\paramsqfr{i}<\deg\elimsqfr$ for $1\le i\le n$.

By construction, the minimal polynomial of $x_n$ in
$\frac{\pr}{I}$ is the same as the minimal polynomial of
$\mulmat$ and is of degree at most $D$. This polynomial is also called
the \emph{eliminating polynomial of $x_n$} and
is
$\elim$.
\revision{Let $\vvec_0\in\field^D$ be
  a column-vector chosen at random and for all $1\leq k<2 D$,
  $\vvec_k^t=\vvec_0^t \mulmat^k$. Assume that the monomial $1$ is the
  first one in the monomial basis $\mbasis$. Then, using Wiedemann's algorithm,
  $\elim$ is computed by 
  guessing the minimal recurrence relations of the table
  $(u_k)_{0\leq k<2 D}$ defined by $u_k=v_{k,1}$.
}
This guessing step
is usually done with the
Berlekamp--Massey algorithm and its fast variants~\cite{BrentGY1980}.

Assuming $x_i$ is the $j$th monomial in $\mbasis$, its parametrization
$\param{i}$ is
computed by solving a Hankel system with matrix
$(\vvec_{k+\ell,1})_{0\leq k,\ell<D}$
and vector $(\vvec_{k,j})_{0\leq k<D}$.
Then, $\paramsqfr{i} = \param{i} \bmod \elimsqfr$.

Whenever the ideal $I$ does not satisfy assumption~\assumesl, a parametrization of
the solutions might still be possible. This is the case if $\sqrt{I}$
satisfies \assumesl.
In that case,
%
for $1\leq i\leq n$, assuming $x_i$ is the $j$th monomial in
$\mbasis$,
$\paramsqfr{i}$ can be computed in
a similar fashion using~\cite[Algorithm~2]{Hyun2020163}. Let
$d=\deg\elim<D$ and
$h=\elim\sum_{k=0}^{d-1}\vvec_{k,1} x_n^{d-1-k} \quo x_n^d$.
Then, 
\begin{equation}
\paramsqfr{i}=
  -\left(\elim\sum_{k=0}^{d-1} \vvec_{k,j} x_n^{d-1-k} \quo x_n^d\right)
  h^{-1} \bmod \elimsqfr.\label{eq:paramsqfr}
\end{equation}

This yields the following algorithm in pseudo-code.
\begin{algorithm}
\caption{Sparse \fglm} \label{alg:sparse_fglm}
\begin{algorithmic}[1]
    \Require {$G \in \field[x_1,\ldots,x_n]$ the reduced \gb for a
      zero-dimensional ideal of degree $D$, satisfying
      assumption~\assumesc
      \wrt
      \revision{$\sdrl$, such that for all $i$ $x_i\not\in L_{\sdrl}(G)$,
        and whose radical
        satisfies assumption~\assumesl}}
    \Ensure {A parametrization of the roots of $\left<G\right>$.}
    \State{Build $\mulmat$ the multiplication matrix of $x_n$}
    \State{Pick $\vvec_0\in\field^D$ at random}
    \For{$k$ from $0$ to $2 D-2$}
        \State{$\vvec_{k+1}\gets \mulmat^t \vvec_k$}
    \EndFor
    \State{$\elim\gets\Berlekamp(V_{0,1},\ldots,V_{2 D-1,1})$}
    \State{$\elimsqfr\gets\squarefree(\elim)$}
    \If{$\deg\elim=D$}
       \For{$i$ from $1$ to $n-1$}
           \State{Solve the Hankel system to determine
             $\param{i}(x_n)$}
           \State{$\paramsqfr{i}\gets\param{i}\bmod\elimsqfr$}
       \EndFor
    \Else
        \For{$i$ from $1$ to $n-1$}
        \State{Compute $\paramsqfr{i}$ as in Equation~\eqref{eq:paramsqfr}.}
        \EndFor
    \EndIf
    \State{return $\{\elim,\elimsqfr,x_{n-1}+\paramsqfr{n-1}(x_n),\ldots,
      x_1+\paramsqfr{1}(x_1)\}$.}
\end{algorithmic}
\end{algorithm}

\revision{Note that for solving purpose, assumptions~\assumesc and~\assumesl can
  always be retrieved by adding to the input system a generic linear form
  depending on one more variable which will then stand as the least one.}

\vspace*{-1em}
\subsection{Univariate Polynomial Solving}
\label{subsec:usolve}

In this subsection, we describe the algorithm used for real root isolation 
in \msolve. It takes as input $\up \in \Q[x]$ which we assume to be
squarefree since the algebraic representation output by \msolve stands for the 
\emph{radical} of the ideal generated by the input equations.  

Hence, let $\up \in \Q[x]$ be squarefree; further, we denote by $\sigma(\up)$ 
the number of sign variations in the sequence of coefficients of $\up$ when 
it is encoded in the standard monomial basis. Note that when $\sigma(\up) = 
0$, $\up$ has no positive real root. By Descartes' rule of signs, the 
difference between $\sigma(\up)$ and the number of positive real roots of 
$\up$ is an even non-negative integer which we denote by $\delta(\up)$.  
Consequently, when $\sigma(\up) = 1$,  $\up$ has a single positive real 
root.

This can be used efficiently in a subdivision scheme, introduced by Akritas and
Collins in~\cite{CoAk76}, as follows. We start by computing an integer $B$ such
that all positive real roots of $\up$ lie in the interval $]0, B[$ using the
bounds given in \eg~\cite[Chapter 10]{BaPoRo}. Note that, up to scaling, one can
assume this interval to be $]0, 1[$. Hence, the idea is to apply some
transformation  $\tilde{\up} = (x+1)^{\deg(\up)}f\left (\frac{1}{x+1}\right )$, and compute $\sigma(\tilde{\up})$. If
it is $0$ or $1$, we are done. Else, one performs recursive calls to the
algorithm by splitting the interval $]0, 1[$ to $\left ]0, \frac{1}{2}\right [$
and $\left ]\frac{1}{2}, 1\right [$. This is done by mapping them to $]0, 1[$,
applying the transformations $x\to \frac{x}{2}$ and $x\to \frac{x+1}{2}$ to
$\tilde{\up}$ and taking the numerator. Termination of this subdivision scheme
is ensured by Vincent's theorem~\cite{Vincent}.


To get all the real roots of $f$ it suffices to apply the transformation $x \to
-x$ and call the subdivision scheme on this newly obtained polynomial. Many
improvements have been brought during the past years, in particular by
integrating Newton's method to accelerate the convergence of the subdivision
scheme (see \eg~\cite{MeSa16}).


\vspace*{-1em}
\section{Implementational Details}
\label{sec:implementation}
In this section, we are given $F\subset \pr$ and we denote by $I$ the ideal 
generated by  $F$. We assume that the base field $\field$ is either  $\Q$ 
or a prime field of characteristic  $ < 2^{31}$. 

To tackle systems with coefficients in  $\Q$, we use multi-modular 
approaches. Here, we do not discuss details on technical 
necessities like good or bad primes in detail, but refer to~\cite{arnold-2003, 
traverso-trace-1988}. Our implementations of \ffour, the linear algebra routine on 
which it relies and \fglm run over prime fields with characteristic $ < 
2^{31}$. In the end, we obtain rational parametrizations with polynomials 
with coefficients in $\Z$. The real root isolator implemented in \msolve is 
based on the big num \texttt{mpz} arithmetic of \gmp~\cite{gmp}.

\vspace*{-0.5em}
\subsection{Efficiency in \ffour}
\label{subsec:efficiency-in-f4}
For an efficient implementation of \ffour we use different approaches.

(1) We use hashing tables with linear probing in order to store the 
exponent vectors
        corresponding to monomials.
    
(2) For testing monomial divisibility in the symbolic preprocessing step
        we use a divisor mask of $32$-bits, if there are more than $32$
        variables we just recognize the first
        $32$.

(3) In general, rows are stored in a sparse format since for most
        systems \ffour matrices are very sparse. For denser matrices a
        sparse-dense hybrid format is implemented.
    
(4) We use the sparsest possible rows as pivot rows when applying Gaussian
        Elimination.

        (5) For computations modulo prime numbers $2^{30} < p < 2^{31}$ we can
        use CPU intrinsics to make the basic operations, additions and
        multiplications of \texttt{uint32\_t} elements more efficient. Using
        \texttt{AVX2} we can store eight $32$-bit (unsigned) coefficients in one
        \mbox{$256$-bit} \texttt{\_\_m256i} type. We apply four multiplications
        and subtractions at a time storing intermediate results in $64$-bit
        (signed) integers. Testing if the intermediate values are negative we
        can add, in that instance, a square of the field characteristic to
        correct positive coefficients of the usual storage type.
        \revision{Depending on the sparsity of the matrix this approach can
          lower the time spent for linear algebra in \ffour by more than the
          half.}

\vspace*{-0.5em}
\subsection{Probabilistic Linear Algebra}
\label{sec:probabilistic-la}
For
\ffour
we use the Gebauer--M\"oller installation
from~\cite{gmInstallation1988} in order to discard useless critical pairs. Since
there still might be zero reductions during the run of the algorithm we apply
over finite fields an idea that was first publicly stated by Monagan and Pearce
in~\cite{monagan-pearce-pasco-2017} (where it is attributed to A.\ Steel 
from the \magma team).

After having moved the sparsest row for each pivot into the upper pivot matrix part,
we take the remaining $k$ rows into account. These are the rows to be reduced by
the upper pivot matrix (\ie the known leading terms for $G$). We partition
these $k$ rows into blocks of a given size, say $\ell$ rows form one block. Now
we take a random linear combination of these $\ell$ rows and reduce it \wrt
the upper pivot matrix. If the outcome is non-zero we have found a new pivot
row and add it to the upper pivot row. Then we take another random linear
combination of the $\ell$ rows. We stop with the current block once we have
either reduced $\ell$ linear combinations or once the first reduction to zero
happens. The probability of getting zero by chance is roughly $1/p$, for $p$
being the field characteristic. So, if $p$ is big enough we get the correct
result with a high probability. Moreover, one can increase the probability of
correctness by doing more than one reduction to zero before the block is
finished.
Once all blocks are handled, we are done with the linear algebra part of \ffour.
Further we call this
strategy
\textbf{probabilistic linear algebra}
.

\vspace*{-0.5em}
\subsection{\ffour Tracer}
In order to have more efficient modular runs of \ffour we can exploit 
already
known meta data from previous runs.
We \textbf{learn} from the first finite field computation modulo some given
prime number $p$ applying \ffour with \emph{exact} linear algebra:
\textbf{Trace} the main steps of the algorithm, \ie for the first round of \ffour
\begin{enumerate}
\item store all polynomials and multiples that generate the matrix,
\item remove from this list all polynomials that are reduced to zero; also
remove all reducers that are only needed for these specific polynomials.
\end{enumerate}
In the following calls of \ffour for different prime numbers we \textbf{apply} the trace
from the computation modulo $p$. For each round we just run the following two
steps:\\
$(1)$ Generate the matrix with the already computed polynomials using the
information from the trace.\\
$(2)$ Use exact linear algebra, add the new polynomials to the basis.

\begin{remark}
\label{rem:tracer}
If we use the tracer to \ffour we cannot use the probabilistic linear
algebra in the first round since then we could not detect which specific rows
reduce to zero. In the application phase of the tracer it is then useless to apply
the probabilistic linear algebra since the matrices are already optimal in the sense
that we do not compute any zero reduction at all. If the first prime number for which
we generate the tracer is a good prime number we can be sure that only a finite
number of other prime numbers exist such that the \gb computed modulo these
primes via applying the tracer is not correct. 
\end{remark}

\vspace*{-1em}
\subsection{Change of orders}

Recall that we apply \fglm to the \emph{generic} situation where the ideal $I$ satisfies 
assumption~\assumesl and that the monomial basis $\mbasis = (m_1, \ldots,\penalty-1000 
m_D)$ of $\frac{\pr}{I}$ ($D$ is the degree of $I$)  satisfies
assumption~\assumesc. This is deduced from the reduced $\sdrl$ \gb $G$ of $I$. 
We denote by $\mulmat$ the matrix encoding the endomorphism $\varphi: 
\overline{f}\in \frac{\pr}{I}\to \overline{f}x_n\in \frac{\pr}{I}$. 

The algorithm in~\cite{sparse-FGLM} relies on computing the Krylov 
sequence:
\[
    \vvec_i^t = \vvec_0^t \ \mulmat^i\ \text{ for } 1\le i < 2D
\]
where $\vvec_0$ is a randomly chosen vector with coefficients in our base  
field $\field$ (which is prime of characteristic ${}< 2^{31}$ in our context).

In~\cite{sparse-FGLM}, Faug\`ere and Mou note that, under
assumption~\assumesc, the matrix $\mulmat$ can be read on the \sdrl \gb of $I$ as 
follows.
\begin{itemize}
    \item[(1)] if $\varphi(m_i) = m_j\in \mbasis$ then the $i$th column of  
        $\mulmat$ is the vector whose entries are all  $0$ except the  
        $j$th which is $1$;
    \item[(2)] if $\varphi(m_i)$ is the lead monomial of the  $j$th 
        element $g_j$ of $G$ then the  $i$th column of $\mulmat$ 
        is the vector of coefficients of the tail of $-g_j$ which is
        $\lt{g_j}{}-g_j$. 
\end{itemize}
In the end, observe that the transpose of $\mulmat$ enjoys a structure of 
generalized companion matrix with "trivial" blocks (corresponding to case 
(1)) and "dense" lines (corresponding to case (2)) which leads to see this 
matrix as a "sparse" one.

In~\cite{sparse-FGLM}, the authors analyze the sparsity of $\mulmat$ under 
some genericity assumptions. In~\cite{Hyun2020163}, the authors develop 
block Krylov techniques to accelerate these algorithms in particular 
through parallelism and make clearer how to apply them in the situation 
where $I$ is not radical. The implementation developed there is based on 
the 
\href{http://eigen.tuxfamily.org/index.php?title=Main_Page}{\texttt{eigen}} 
library for sparse matrix multiplication.
In our implementation, we treat $\mulmat$, not as a general sparse matrix 
but as a generalized companion matrix. We encode the transpose of $\mulmat$ 
as follows:
 \begin{itemize}
    \item we store the position of "trivial" lines and, for these lines, 
        the position of the '1' in these "trivial" lines;
    \item we store the position of "dense" lines and an array for the list 
        of coefficients. 
\end{itemize}
With such an encoding, computing the $\vvec_i$'s simply boils down to 
multiplying the "dense" rows of  $\mulmat$ with a subvector of  $\vvec_{i-1}$ 
and copying entries of $\vvec_{i-1}$ to the appropriate coordinates of $\vvec_i$.  

This reduction to dense matrix vector multiplication is efficient if most of the
"dense" lines are indeed dense which is the case in most of the examples. It
also allows us to use in a straightforward way \texttt{AVX2} intrinsics for
computing scalar products of vectors with coefficients in finite fields. As
explained in Subsection~\ref{subsec:efficiency-in-f4}, we can then perform four
multiplications of the entries of our vectors by storing them in a
\texttt{\_\_m256i} register. To delay as much as possible reductions by the
prime number defining our base field, we accumulate the highest and lowest $32$
bits in separate accumulators. \revision{Since we are dealing with dense
  vectors, this approach allows us to obtain a speed-up close to $3$. }

\subsubsection*{Verifying the Parametrizations}
If $\deg\elim = D$,
then the returned \gb is the reduced one of $I$ for
$\slex$. Otherwise, if $\sqrt{I}$ satisfies assumption~\assumesl,
the goal is to return a \gb of this ideal.
We describe now how we implemented a new procedure deciding if $\sqrt{I}$ 
satisfies \assumesl.


\revision{In Section~\ref{subsec:fglm}, we computed a polynomial
$p_i=x_i+\paramsqfr{i}(x_n)$ as the parametrization of $x_i$ in $\sqrt{I}$.
We now compute a second parametrization $q_i$ for $x_i$ and
compare them.} Since computing the sequence terms is
actually the bottleneck of this variant of the \fglm algorithm, the
goal is to use the sequence terms at hand.

Let us notice that since $(\vvec_k)_{k\geq 0}$ satisfies the relation given
by $\elim$,
it is hopeless to
just shift the sequence terms by increasing $k$.
Using
this recurrence relation,
we can rewrite the computations \wrt the first sequence terms making them 
yield
\revision{$q_i=p_i$, whether $\sqrt{I}$ satisfies \assumesl or
  not.}
Therefore, the idea is to shift the
sequence terms in \emph{another direction}. \revision{Let us assume
  that $1$
  (\resp $x_i$, \resp $x_i^2$) is the
  first (\resp $j$th, \resp $j'$th) monomial in $\mbasis$ and let pick
  $\lambda\in\field$ at random. Replacing all instances of
  $\vvec_{k,1}$ by $\vvec_{k,j}+\lambda \vvec_{k,1}$ and all
  those of $\vvec_{k,j}$ by $\vvec_{k,j'} + \lambda\vvec_{k,j}$
  makes us
  compute a parametrization $q_i=x_i+\tilde{\param{}}_i(x_n)$ of
  the radical of the colon ideal $I\,\colon (x_i+\lambda)$,
  see~\cite[Th.~3.1]{colonideal}. If $\field$ is large enough,
  then $I\,\colon (x_i+\lambda)=I$ and both ideals share the same
  radical. Now, if $\sqrt{I}$ satisfies assumption~\assumesl, then so
  does $\sqrt{I\,\colon (x_i+\lambda)}$. Otherwise,
  $q_i$ actually depends on $\lambda$ and must be
  different from $p_i$, the computed parametrization of $x_i$ for
  $\sqrt{I}$. Thus, we know
  that $\sqrt{I}$ does not satisfy assumption~\assumesl.}

\subsection{Multi-modular Approach}
When $\pr=\Q[x_1, \ldots, x_n]$, we have implemented efficient 
multi-modular algorithms. 
%
%

One starts by picking randomly a prime number $p_0$ in the interval 
$]2^{30},2^{31}[$ 
and next \emph{(i)} run the $\ffour$ tracer on the modular image of our input
system in $\frac{\Z}{p_0\Z}[x_1, \ldots, x_n]$, \emph{(ii)} run \fglm on the
computed \gb and normalize the obtained \gb for $\slex$ (which is in Shape
position by assumption) to obtain a rational parametrization. This process is
repeated for several primes, applying the tracer we learnt from $p_0$
until one can perform rational reconstruction (through Chinese remainder 
lifting) to obtain a solution over $\pr$ whose modular image by reduction 
to some prime $p$ coincides with the output of step \emph{(ii)} when 
running the computation over $\frac{\Z}{p\Z}[x_1, \ldots, x_n]$.
For Chinese remainder lifting and rational reconstruction, we use functions 
from \texttt{FLINT}~\cite{flint} (which we have slightly adapted to our context). 

Note that in step \emph{(i)}, one can replace the \ffour tracer with \ffour 
based on probabilistic linear algebra. Note also that all computations 
modulo prime numbers are \emph{independent} of each other.

This multi-modular approach is probabilistic: only for homogeneous systems we
can apply a final check (over \Q) if the computed \gb is
correct. 
Other than that we get the correct result if the \gb computed 
modulo the first chosen prime $p_0$ coincides with the image modulo $p_0$ 
of the \gb (over the rationals) of the input system. This happens with high 
probability and the number of such bad primes is finite (see 
\eg~\cite{traverso-trace-1988, bad-primes-paper-2016}).  

One choice in the current design of \msolve, which is inspired by the last
release of \fgb, is that the multi-modular 
process is implemented \emph{globally}, \ie we do not lift the 
intermediate reduced \sdrl \gb over $\Q$.

\subsection{Univariate real root isolation}

Our implementation uses tricks which were previously introduced by Hanrot 
et al.\ in \url{https://members.loria.fr/PZimmermann/software/} to implement 
\cite{RoZi04} and also used in the \texttt{SLV} library \cite{TsigSLV}.  
These consist in observing that we only need the two basic operations: 
\emph{(i)} shifting $x \to  x + 1$ in the considered polynomial and 
\emph{(ii)} scaling the coefficients by the transformation $x \to 2^k x$ 
for $k \in \Z$ which can be handled by specific
\gmp
\texttt{mpz\_} shift operators~\cite{gmp}.  The single innovation in \msolve is 
motivated by
the large bit sizes of the coefficients (several tens of thousands) and
the large degrees (several thousands)
of the polynomials output by \msolve. 

Firstly, observe that one needs to count the number of sign variations of 
the polynomial obtained after a combination of \emph{(i)} and \emph{(ii)}.  
In our context the bit size of the coefficients is way larger than the 
degree of the considered polynomial.  Hence, taking appropriate dyadic 
approximations of these coefficients is
sufficient to decide the sign (unless some unexpected cancellations occur).  
Note that computing such dyadic approximations is free using \texttt{GMP}. 

Secondly, to tackle large degrees, we revisit asymptotically fast 
algorithms for Taylor shift (see~\cite{GaGe97}) (which we combine with the 
above dyadic approximation technique) and implement them carefully using 
the \texttt{FFT}-based multiplication of \texttt{FLINT} for univariate 
polynomials with integer coefficients.  This is a major difference with
other implementations because of the (wrong) belief that asymptotically 
fast algorithms are useless in this context
(see~\cite[Section~3.1]{KoRoSa16}).  This allows us to obtain a univariate solver which 
outperforms the state of the art on examples coming from our computations 
(usually extracted from applications of polynomial system solving). The 
cross-over point of our asymptotically fast implementation of the Taylor 
shift against the classical implementations used in current real root 
solvers is around degree $512$. 

Similarly, we implement the quadratic interval real root refinement 
described in~\cite{Abbott14} for better practical efficiency which improves 
upon the naive one implemented in \texttt{SLV}.



\section{Experimental Results}
\label{sec:results}
We compare \msolve with two other computer algebra systems:
\begin{itemize}
\item \magma\textsc{-v2.23-6}~\cite{magma}:
  using
  the command \texttt{Variety()}.
\item \maple\textsc{-v2019}~\cite{maple2019}:
  using
  the command \texttt{PolynomialSystem()} from the module
  \texttt{SolveTools} with option \texttt{engine=groebner}.
\end{itemize}
All compared implementations use Faug\`ere's \ffour algorithm and variants of
the \fglm algorithm and then solve univariate polynomials.


All chosen systems are zero-dimensional with rational coefficients. All
computations are done sequentially.  Table~\ref{table:first-example} states various, partly well-known
benchmarks, which differ in their specific hardness, like reduction process,
pair handling, sparsity of multiplication matrices, etc.
Table~\ref{table:critical-points-example} is dedicated to critical
points computations, \texttt{CP(d,nv,np)}
describes critical points for a system of \texttt{np} polynomials in \texttt{nv}
variables of degree \texttt{d}. 
{\small
\begin{table*}[h]
  \centering
  \renewcommand{\tabcolsep}{1mm}
  \def\arraystretch{1.1}
  \scalebox{1}{
      \begin{tabular}{|c||r|c||r|r|r|r||r|r|r||>{\raggedleft\arraybackslash}p{4em}|r||r|r|}
    \hline
    \multirow{2}{*}{Examples} &
    \multicolumn{2}{c||}{System data} &
    \multicolumn{4}{c||}{\msolve single modular computation} &
    \multicolumn{3}{c||}{\msolve overall} &
    \multicolumn{2}{c||}{\revision{\maple single modular}} &
    \multicolumn{2}{c|}{Others overall}\\
    &
    \multicolumn{1}{c|}{degree} &
    \multicolumn{1}{c||}{radical} &
    \multicolumn{1}{c|}{\ffour (prob.)} &
    \multicolumn{1}{c|}{\ffour (learn)} &
    \multicolumn{1}{c|}{\ffour (apply)} &
    \multicolumn{1}{c||}{\fglm} &
    \multicolumn{1}{c|}{\# primes} &
    \multicolumn{1}{c|}{trace} &
    \multicolumn{1}{c||}{independent} &
    \multicolumn{1}{c|}{\ffour} &
    \multicolumn{1}{c||}{\fglm} &
    \multicolumn{1}{c|}{\maple} &
    \multicolumn{1}{c|}{\magma}\\
    \hline
    \href{https://gitlab.lip6.fr/eder/msolve-examples/-/raw/master/zero-dimensional/kat9-qq.ms}{Katsura-9}
    & $256$ & yes & $0.06$ & $0.17$ & $0.03$ & $0.03$ & $83$ &
    $\textcolor{red!80!black}{4.89}$ & $7.49$ & $0.10$ & $0.04$ & $104$ & $2,522$ \\
    \href{https://gitlab.lip6.fr/eder/msolve-examples/-/raw/master/zero-dimensional/kat10-qq.ms}{Katsura-10} & $512$ & yes & $0.24$ & $0.81$ & $0.09$ & $0.11$ & $188$ &
    $\textcolor{red!80!black}{43.7}$ & $70.5$ & $0.36$ & $0.15$ & $1,278$ & $82,540$ \\
    \href{https://gitlab.lip6.fr/eder/msolve-examples/-/raw/master/zero-dimensional/kat11-qq.ms}{Katsura-11} & $1,024$ & yes & $1.34$ & $6.26$ & $0.45$ & $0.49$ & $388$ &
    $\textcolor{red!80!black}{424}$ & $814$ & $1.82$ & $0.74$ & $7,812$ & $-$ 
      \\
    \href{https://gitlab.lip6.fr/eder/msolve-examples/-/raw/master/zero-dimensional/kat12-qq.ms}{Katsura-12} & $2,048$ & yes & $8.61$ & $56.1$ & $3.10$ & $3.96$ & $835$ &
    $\textcolor{red!80!black}{6,262}$ & $11,215$ & $8.50$ & $5.40$ & $120,804$ & $-$ \\
    \href{https://gitlab.lip6.fr/eder/msolve-examples/-/raw/master/zero-dimensional/kat13-qq.ms}{Katsura-13} & $4,096$ & yes & $52.8$ & $425$ & $18.9$ & $30.6$ & $1,772$ &
    $\textcolor{red!80!black}{89,390}$ & $148,372$ & $60.9$ & $35.7$ & $-$
                                & $-$ \\
    \href{https://gitlab.lip6.fr/eder/msolve-examples/-/raw/master/zero-dimensional/kat14-qq.ms}{Katsura-14} & $8,192$ & yes & $318$ & $3,336$ & $128$ & $210$ & $3,847$ &
    $\textcolor{red!80!black}{1,308,602}$ & $2,007,170$ & $393$ & $271$ & $-$ & $-$\\[0.2em]
    \href{https://gitlab.lip6.fr/eder/msolve-examples/-/raw/master/zero-dimensional/eco10-qq.ms}{Eco-10} & $256$ & yes & $0.10$ & $0.28$ & $0.05$ & $0.02$ & $161$ &
    $\textcolor{red!80!black}{12.5}$ & $21.2$ & $0.14$ & $0.03$ & $26.3$ & $6,520$ \\
    \href{https://gitlab.lip6.fr/eder/msolve-examples/-/raw/master/zero-dimensional/eco11-qq.ms}{Eco-11} & $512$ & yes & $0.39$ & $1.21$ & $0.17$ & $0.07$ & $327$ &
    $\textcolor{red!80!black}{90.3}$ & $161$ & $0.56$ & $0.12$ & $312$ & $214,770$ \\
    \href{https://gitlab.lip6.fr/eder/msolve-examples/-/raw/master/zero-dimensional/eco12-qq.ms}{Eco-12} & $1,024$ & yes & $2.25$ & $11,619$ & $1.07$ & $0.34$ & $530$ &
    $\textcolor{red!80!black}{877}$ & $1,619$ & $2.97$ & $0.85$ & $4,287$ & $-$ \\
    \href{https://gitlab.lip6.fr/eder/msolve-examples/-/raw/master/zero-dimensional/eco13-qq.ms}{Eco-13} & $2,048$ & yes & $11.7$ & $67.3$ & $6.61$ & $2.12$ & $1,225$ &
    $\textcolor{red!80!black}{12,137}$ & $19,553$ & $15.1$ & $6.70$ & $66,115$ & $-$ \\
    \href{https://gitlab.lip6.fr/eder/msolve-examples/-/raw/master/zero-dimensional/eco14-qq.ms}{Eco-14} & $4,096$ & yes & $67.1$ & $516$ & $34.8$ & $25.9$ & $2,670$ &
    $\textcolor{red!80!black}{167,798}$ & $254,389$ & $104.8$ & $69.1$ & $-$
                                & $-$ \\[0.2em]
    \href{https://gitlab.lip6.fr/eder/msolve-examples/-/raw/master/zero-dimensional/henrion5.ms}{Henrion-5} & $100$ & yes & $0.01$ & $0.01$ & $0.004$ & $0.01$ & $83$ &
    $\textcolor{red!80!black}{0.71}$ & $0.83$ & $0.01$ & $0.01$& $2.7$ & $93$ \\
    \href{https://gitlab.lip6.fr/eder/msolve-examples/-/raw/master/zero-dimensional/henrion6.ms}{Henrion-6} & $720$ & yes & $0.11$& $0.22$ & $0.07$ & $0.11$ & $612$ &
    $\textcolor{red!80!black}{138}$ & $157$ & $0.17$ & $0.16$ & $1,470$ &
    $-$
      \\
    \href{https://gitlab.lip6.fr/eder/msolve-examples/-/raw/master/zero-dimensional/henrion7.ms}{Henrion-7} & $5,040$ & yes & $9.55$ & $27.5$ & $6.51$ & $20.46$ & $4,243$ &
    $\textcolor{red!80!black}{117,803}$ &
    $127,456$ & $12.8$ & $27.1$ & $-$
                                & $-$ \\[0.2em]
    \href{https://gitlab.lip6.fr/eder/msolve-examples/-/raw/master/zero-dimensional/noon7-qq.ms}{Noon-7} & $2,173$ & yes & $1.66$ & $5.3$& $0.93$ & $1.95$ & $1,305$ &
    $4,039$ & $5,045$ & $1.97$ & $3.13$ & $\textcolor{red!80!black}{432}$ & $-$
      \\
    \href{https://gitlab.lip6.fr/eder/msolve-examples/-/raw/master/zero-dimensional/noon8-qq.ms}{Noon-8} & $6,545$ & yes & $26.6$ & $153$ & $17.5$ & $72.3$ & $6,462$ &
    $598,647$ & $640,177$ & $32.4$ & $76.2$& $\textcolor{red!80!black}{5,997}$ & $-$
      \\[0.2em]
    \href{https://gitlab.lip6.fr/eder/msolve-examples/-/raw/master/zero-dimensional/phuoc1.ms}{Phuoc-1} & $1,102$ & no & $4.01$ & $4.65$ & $3.42$ & $2.59$ & $753$ &
    $\textcolor{red!80!black}{4,467}$ & $5,056$ & $4.60$& $5.91$& $-$ 
                                & $-$ \\
    \hline
    \end{tabular}
}
  \caption{Benchmark timings given in seconds (if not otherwise stated)}
  \label{table:first-example}\vspace{-0.5cm}
\end{table*}
}

For each system we give its degree and if it is radical \revision{(all but one
  are radical)}. For \msolve we give
specific timing information, also on the single modular computations: We apply
\msolve with the \texttt{tracer} option, giving also the timings for the first
modular computation learning and generating the tracer (\texttt{F4\,(learn)}) and
the timings for the further modular computations applying only the tracer
(\texttt{F4\,(apply)}). We also use \msolve with \texttt{independent} modular
computations, applying the probabilistic linear algebra in each modular \ffour
(\texttt{F4\,(prob.)}) In any case, we apply the same \fglm implementation.
Furthermore, we state the number of primes needed by \msolve to solve over $\Q$.
For \maple and \magma we just give the overall timings. Symbol '$-$' means that
the computation was stopped after waiting more than $10$ times the runtime of
\msolve. \revision{For all systems, the bottleneck has been the computation of
  either a Lexicographical \gb in shape position or a rational parametrization
  of the solution set}. 

First thing to note is that \magma is in all instances slower than \msolve or
\maple. Although, for some
examples, \magma's modular \ffour computation is even a bit faster than the
other two, \magma's bottleneck is \revision{both} a not optimized \fglm
\revision{combined with the fact that \magma seems to lift a lexicographical \gb
  instead of a rational parametrization (the latter one having in general
  coefficients of significantly smaller bit size)}. 

For nearly all systems, \msolve is faster, sometimes by an order of magnitude,
than \maple. \revision{We report on modular timings of \maple for \ffour (which
  is based on a probabilistic linear algebra) and \fglm. It appears that
  \msolve's modular implementations of both \ffour and \fglm are faster than the
  ones in \maple (with a speed-up sometimes close to $2$, sometimes less). It
  seems that in the multi-modular process, \maple uses its probabilistic variant
  of \ffour while \msolve takes advantage of its tracer. Also, \maple's
  documentation indicates that on some examples, an algorithm computing a
  so-called rational univariate representation (preserving multiplicities),
  instead of \fglm, may be used. It is likely that on most examples we tried,
  \fglm is not used. Note also that \maple lifts a whole $\DRL$ \gb over $\Q$
  while \msolve avoids this step. Furthermore, our tracer shares some
  similarities with~\cite{Xcas}.}


There are, of course, few examples, where \msolve is not competitive.
\revision{In particular, for some systems, \msolve may need to introduce a
  generic linear form as previously explained while a rational parametrization
  can be obtained without it (but up to computing normal forms). Also some
  systems admit a triangular representation, and/or can be split. It seems
  that \maple can detect and sometimes take advantage of such situations. This
  is typically the case for the \texttt{Noon-n} examples.}


As for the univariate solver in \msolve, we compare with \maple\footnote{We use
  \maple -v16 as it is faster than the -v2019 for real root isolation on our benchmarks} and
\tdescartes (non-open source) and \slv (open source). We use the standard
benchmarks provided in Table~\ref{table:first-example}, the solving process
leads to polynomials which do not have clusters of real roots. The benefit of
implementing asymptotically fast algorithms for real root solvers is now
obvious: \msolve's runtimes outperforms its competitors' on this class of
problems.  

\revision{Finally, Table~\ref{tab:memory} compares memory usage of \msolve
  against \maple and \magma. It illustrates the low memory usage of \msolve.
  However, we emphasize that \msolve is a specialized library while \maple and
  \magma are general purpose computer algebra systems.}

Overall, \msolve performs very efficiently on a wide range of input systems,
using way less memory than its competitors, allowing its users to solve
polynomial systems which are not tractable by \maple and \magma.

{\small
\begin{table*}[h]
  \centering
  \renewcommand{\tabcolsep}{1mm}
  \def\arraystretch{1.1}
  \scalebox{1}{
    \begin{tabular}{|c||r|c||r|r|r|r||r|r|r||>{\raggedleft\arraybackslash}p{4em}|r||r|r|}
    \hline
    \multirow{2}{*}{Examples} &
    \multicolumn{2}{c||}{System data} &
    \multicolumn{4}{c||}{\msolve single modular computation} &
    \multicolumn{3}{c||}{\msolve overall} &
    \multicolumn{2}{c||}{\revision{\maple single modular}} &
    \multicolumn{2}{c|}{Others overall}\\
    &
    \multicolumn{1}{c|}{degree} &
    \multicolumn{1}{c||}{radical} &
    \multicolumn{1}{c|}{\ffour (prob.)} &
    \multicolumn{1}{c|}{\ffour (learn)} &
    \multicolumn{1}{c|}{\ffour (apply)} &
    \multicolumn{1}{c||}{\fglm} &
    \multicolumn{1}{c|}{\# primes} &
    \multicolumn{1}{c|}{trace} &
    \multicolumn{1}{c||}{independent} &
    \multicolumn{1}{c|}{\ffour} &
    \multicolumn{1}{c||}{\fglm} &
    \multicolumn{1}{c|}{\maple} &
    \multicolumn{1}{c|}{\magma}\\
    \hline
    \href{https://gitlab.lip6.fr/eder/msolve-examples/-/raw/master/zero-dimensional/cp_d_3_n_5_p_2.ms}{CP$(3,5,2)$} & $288$ &yes &  $0.03$ & $0.04$ & $0.01$ & $0.03$ & $326$ &
    $\textcolor{red!80!black}{18.1}$ & $19.2$ & $0.06$ & $0.05$ & $249$ & $-$ 
      \\
    \href{https://gitlab.lip6.fr/eder/msolve-examples/-/raw/master/zero-dimensional/cp_d_3_n_6_p_2.ms}{CP$(3,6,2)$} & $720$& yes & $0.22$ & $0.59$ & $0.12$ & $0.16$ & $1,042$ &
    $\textcolor{red!80!black}{390}$ & $450$ & $0.31$ & $0.22$& $23,440$ & $-$\\
    \href{https://gitlab.lip6.fr/eder/msolve-examples/-/raw/master/zero-dimensional/cp_d_3_n_7_p_2.ms}{CP$(3,7,2)$} & $1,728$ & yes & $1.97$ & $8.18$ & $1.23$& $1.20$ & $3,037$ &
    $\textcolor{red!80!black}{9,643}$ & $11,511$ & $2.78$ & $2.54$ & $-$
                                & $-$\\
    \href{https://gitlab.lip6.fr/eder/msolve-examples/-/raw/master/zero-dimensional/cp_d_3_n_8_p_2.ms}{CP$(3,8,2)$} & $4,032$ & yes & $18.5$ & $111.5$ & $12.2$ & $19.6$ & $8,211$ &
    $\textcolor{red!80!black}{269,766}$ & $323,838$ & $24.6$ & $25.3$ & $-$ & $-$\\[0.2em]
    \href{https://gitlab.lip6.fr/eder/msolve-examples/-/raw/master/zero-dimensional/cp_d_4_n_4_p_3.ms}{CP$(4,4,3)$} & $576$ & yes & $0.04$ & $0.86$ & $0.03$ & $0.07$ & $339$ &
    $\textcolor{red!80!black}{40.9}$ & $41.8$ & $0.08$ & $0.11$ & $916$ & $-$\\
    \href{https://gitlab.lip6.fr/eder/msolve-examples/-/raw/master/zero-dimensional/cp_d_4_n_5_p_3.ms}{CP$(4,5,3)$} & $3,456$ & yes & $3.24$ & $8.60$ & $2.23$ & $4.83$ & $2,747$ &
    $\textcolor{red!80!black}{21,528}$ & $23,559$ & $4.33$ & $9.21$ & $-$ & $-$\\[0.2em]
    \href{https://gitlab.lip6.fr/eder/msolve-examples/-/raw/master/zero-dimensional/cp_d_3_n_6_p_6.ms}{CP$(3,6,6)$} & $729$ & yes & $0.18$ & $0.42$ & $0.11$ & $0.15$ & $779$ &
    $\textcolor{red!80!black}{255}$ & $294$ & $0.30$ & $0.23$ & $-$
                                & $-$\\
    \href{https://gitlab.lip6.fr/eder/msolve-examples/-/raw/master/zero-dimensional/cp_d_4_n_6_p_6.ms}{CP$(4,6,6)$} & $4,096$ & yes & $7.70$ & $25.6$ & $5.44$ & $14.09$ & $3,476$ &
    $\textcolor{red!80!black}{71,472}$ & $77,941$ & $10.2$ & $16.71$ & $-$
                                & $-$\\
    \href{https://gitlab.lip6.fr/eder/msolve-examples/-/raw/master/zero-dimensional/cp_d_3_n_7_p_7.ms}{CP$(3,7,7)$} & $2,187$ & yes & $2.49$ & $8.97$ & $1.58$ & $1.86$ & $2,795$ &
    $\textcolor{red!80!black}{12,412}$ & $14,375$ & $3.27$ & $3.75$& $-$ & $-$\\
    \hline
    \end{tabular}
}
  \caption{Critical points timings given in seconds (if not otherwise stated)}
  \label{table:critical-points-example}\vspace{-0.8cm}
\end{table*}
\begin{table*}[t]
  \def\arraystretch{1.1}
  \centering
\begin{tabular}{|c|r||r|r|r|r|r|r|r|}
    \hline
    \multirow{2}{*}{Examples} & \multirow{2}{*}{$\sharp$ sols} &
\multicolumn{1}{c|}{{\small \msolve}} &  \multicolumn{2}{c|}{\small \maple} & \multicolumn{2}{c|}{\small SLV} & \multicolumn{2}{c|}{tdescartes} \\
&            &  \multicolumn{1}{c|}{time} & \multicolumn{1}{c|}{time} & \multicolumn{1}{c|}{ratio} & \multicolumn{1}{c|}{time} & \multicolumn{1}{c|}{ratio} &
\multicolumn{1}{c|}{time} & \multicolumn{1}{c|}{ratio} \\
    \hline
        \href{https://gitlab.lip6.fr/eder/msolve-examples/-/raw/master/zero-dimensional/kat10-qq.ms}{Katsura-10}
        &  $120$     & \textcolor{red!80!black}{$3.1$}   &  $4.8$   & $1.5$  & $3.8$ & $1.2$ & $20$ & $6.5$\\
    \href{https://gitlab.lip6.fr/eder/msolve-examples/-/raw/master/zero-dimensional/kat11-qq.ms}{Katsura-11}
    &  $216$   & \textcolor{red!80!black}{$27$}     &  $60$   & $2.2$ & $50.5$ & $1.9$ & $156$ &  $5.8$\\
    \href{https://gitlab.lip6.fr/eder/msolve-examples/-/raw/master/zero-dimensional/kat12-qq.ms}{Katsura-12}
    &  $326$   &  \textcolor{red!80!black}{$207$}   &  $656$  & $3.2$  & $555$ & $2.7$ & $2,206$ & $10.6$ \\
    \href{https://gitlab.lip6.fr/eder/msolve-examples/-/raw/master/zero-dimensional/kat13-qq.ms}{Katsura-13}
    &  $582$   & \textcolor{red!80!black}{$2,220$}   & $16,852$ & $7.6$   & $13,651$ & $6.1$ & $22,945$ &
    $10.3$ \\
    \href{https://gitlab.lip6.fr/eder/msolve-examples/-/raw/master/zero-dimensional/kat14-qq.ms}{Katsura-14}
    &$900$  & \textcolor{red!80!black}{$20,149$}    & $250,094$ & $12.4$   & $252,183$ & $12.5$ &
    $384,566$ & $19.1$ \\
    \href{https://gitlab.lip6.fr/eder/msolve-examples/-/raw/master/zero-dimensional/kat15-qq.ms}{Katsura-15}
    & $1,606$  & \textcolor{red!80!black}{$197,048$}& $3,588,835$  & 18.2   & $3,540,480$ & $18.0$ & $5,178,180$ &  $26.3$\\
    \href{https://gitlab.lip6.fr/eder/msolve-examples/-/raw/master/zero-dimensional/kat16-qq.ms}{Katsura-16}
    & $2,543$  & \textcolor{red!80!black}{$1,849,986$} & $-$ &  $-$ & $-$ &$-$ &$-$ & $-$ \\
    \href{https://gitlab.lip6.fr/eder/msolve-examples/-/raw/master/zero-dimensional/kat17-qq.ms}{Katsura-17}
    & $4,428$  & \textcolor{red!80!black}{$16,128,000$} & $-$ & $-$   &$-$  &$-$ &$-$ & $-$ \\
    \hline
\end{tabular}
  \caption{Real root isolation timings given in seconds}
  \label{tab:univ-solver}\vspace{-0.8cm}
\end{table*}
}
{\small
\begin{table*}[t]
  \def\arraystretch{1}
  \centering
\begin{tabular}{|c||r|r|r|||c||r|r|r|}
    \hline
    Examples&\small \msolve & \small \maple & \small \magma & Examples&\small \msolve & \small \maple & \small \magma \\
    \hline
        \href{https://gitlab.lip6.fr/eder/msolve-examples/-/raw/master/zero-dimensional/kat9-qq.ms}{Katsura-9}
        & \textcolor{red!80!black}{$15$}   &  $271$   & $71$ &
        \href{https://gitlab.lip6.fr/eder/msolve-examples/-/raw/master/zero-dimensional/henrion5.ms}{Henrion-5}
        & \textcolor{red!80!black}{$11$}   &  $26$   & $23$ \\
        \href{https://gitlab.lip6.fr/eder/msolve-examples/-/raw/master/zero-dimensional/kat10-qq.ms}{Katsura-10}
        & \textcolor{red!80!black}{$25$}   &  $276$   & $223$ &
        \href{https://gitlab.lip6.fr/eder/msolve-examples/-/raw/master/zero-dimensional/henrion6.ms}{Henrion-6}
        & \textcolor{red!80!black}{$47$}   &  $171$   & $-$ \\
        \href{https://gitlab.lip6.fr/eder/msolve-examples/-/raw/master/zero-dimensional/kat11-qq.ms}{Katsura-11}
        & \textcolor{red!80!black}{$66$}   &  $2,279$   & $-$ &
        \href{https://gitlab.lip6.fr/eder/msolve-examples/-/raw/master/zero-dimensional/henrion7.ms}{Henrion-7}
        & \textcolor{red!80!black}{$3,428$}   &  $-$   & $-$ \\
        \href{https://gitlab.lip6.fr/eder/msolve-examples/-/raw/master/zero-dimensional/kat12-qq.ms}{Katsura-12}
        & \textcolor{red!80!black}{$229$}   &  $1,123$   & $-$ &
        \href{https://gitlab.lip6.fr/eder/msolve-examples/-/raw/master/zero-dimensional/noon7-qq.ms}{Noon-7}
        & \textcolor{red!80!black}{$209$}   &  $419$   & $-$ \\
        \href{https://gitlab.lip6.fr/eder/msolve-examples/-/raw/master/zero-dimensional/kat13-qq.ms}{Katsura-13}
        & \textcolor{red!80!black}{$1,037$}   &  $-$   & $-$ &
        \href{https://gitlab.lip6.fr/eder/msolve-examples/-/raw/master/zero-dimensional/noon8-qq.ms}{Noon-8}
        & \textcolor{red!80!black}{$881$}   &  $1,227$   & $-$ \\
        \href{https://gitlab.lip6.fr/eder/msolve-examples/-/raw/master/zero-dimensional/eco10-qq.ms}{Eco-10}
        & \textcolor{red!80!black}{$27$}   &  $210$   & $213$ &
        \href{https://gitlab.lip6.fr/eder/msolve-examples/-/raw/master/zero-dimensional/cp_d_3_n_5_p_2.ms}{CP$(3,5,2)$}
        & \textcolor{red!80!black}{$17$}   &  $525$   & $-$ \\
        \href{https://gitlab.lip6.fr/eder/msolve-examples/-/raw/master/zero-dimensional/eco11-qq.ms}{Eco-11}
        & \textcolor{red!80!black}{$82$}   &  $428$   & $354$ &
        \href{https://gitlab.lip6.fr/eder/msolve-examples/-/raw/master/zero-dimensional/cp_d_3_n_6_p_2.ms}{CP$(3,6,2)$}
        & \textcolor{red!80!black}{$55$}   &  $7,885$   & $-$ \\
        \href{https://gitlab.lip6.fr/eder/msolve-examples/-/raw/master/zero-dimensional/eco12-qq.ms}{Eco-12}
        & \textcolor{red!80!black}{$117$}   &  $1,027$   & $-$ &
        \href{https://gitlab.lip6.fr/eder/msolve-examples/-/raw/master/zero-dimensional/cp_d_3_n_7_p_2.ms}{CP$(3,7,2)$}
        & \textcolor{red!80!black}{$312$}   &  $-$   & $-$ \\
        \href{https://gitlab.lip6.fr/eder/msolve-examples/-/raw/master/zero-dimensional/eco13-qq.ms}{Eco-13}
        & \textcolor{red!80!black}{$318$}   &  $8,654$   & $-$ &
        \href{https://gitlab.lip6.fr/eder/msolve-examples/-/raw/master/zero-dimensional/cp_d_4_n_4_p_3.ms}{CP$(4,4,3)$}
        & \textcolor{red!80!black}{$24$}   &  $635$   & $-$ \\
        \href{https://gitlab.lip6.fr/eder/msolve-examples/-/raw/master/zero-dimensional/eco14-qq.ms}{Eco-14}
        & \textcolor{red!80!black}{$15,748$}   &  $-$   & $-$ &
        \href{https://gitlab.lip6.fr/eder/msolve-examples/-/raw/master/zero-dimensional/cp_d_4_n_5_p_3.ms}{CP$(4,5,3)$}
        & \textcolor{red!80!black}{$2,065$}   &  $-$   & $-$ \\
        \href{https://gitlab.lip6.fr/eder/msolve-examples/-/raw/master/zero-dimensional/phuoc1.ms}{Phuoc-1}
        & \textcolor{red!80!black}{$176$}   &  $-$   & $-$ &&&&\\
    \hline
\end{tabular}
  \caption{\revision{Maximal memory usage given in MB}}
  \label{tab:memory}\vspace{-0.8cm}
\end{table*}
}
%
%
%
%
%
%
%
%
%


\vspace{-0.2cm}
\paragraph*{Acknowledgments.} We thank J.-Ch.~Faug\`ere for his advices
and support and for providing us the rational parametrizations of Katsura-n for
$15\leq n\leq 17$, and A.~Bostan for his comments on this paper.

\bibliographystyle{abbrv} \bibliography{msolve}

\end{document}